\documentclass[twocolumn,preprintnumbers,amsmath,amssymb,floatfix,superscriptaddress]{revtex4}

\usepackage{graphicx}
\usepackage{amsmath}
\usepackage{amssymb}
\usepackage{bm}
\usepackage{mathrsfs}
\usepackage{color}
\usepackage[normalem]{ulem}
    \usepackage{braket}

\newcommand{\oad}{\hat{a}^\dagger}
\newcommand{\oa}{\hat{a}}

\newcommand{\oald}{\hat{\alpha}^\dagger}
\newcommand{\oal}{\hat{\alpha}}

\begin{document}

\title{Exact quantization of superconducting circuits}
\author{M.H. Ansari}
\affiliation{ Peter Gr\"unberg Institute, Forschingszentrum J\"ulich, Germany}
\affiliation{ J\"ulich-Aachen Research Alliance (JARA), Fundamentals of Future Information Technologies, Germany}

\begin{abstract}
We present a  theoretical description for circuits consisting of weak anharmonic qubits coupled to cavity multimodes. We obtain a unitary transformation that diagonalizes  harmonic sector of the circuit. Weak anharmonicity does not alter the normal mode basis, however it can modify energy levels.  We study two examples of a transmon  and two transmons coupled to bus resonator, and we determine dressed frequencies and Kerr nonlinearities in closed form formulas. Our results are valid for arbitrary frequency detuning and coupling within and beyond dispersive regime.   
\end{abstract}

\pacs{05.30.-d; 03.67.-a; 03.67.Mn,42.50.Ct}

\maketitle


Quantum computing is rapidly progressing toward practical technology \cite{{Lucero},{Kandala17},{Cai},{Monz}}. One of the leading architectures for a quantum processor is superconducting circuit made of Josephson junction (JJ) nonlinear oscillators \cite{Wilhelm}.  JJ's make the quantum analog of classical bits, i.e. qubits. They are driven by microwave pulses \cite{Chow} and inter-qubit interactions are made possible by coupling individual qubits  to a common  microwave cavity \cite{Niemczyk}. 
State of the art qubit circuits have up to tens of qubits and the key milestone of next few years  research is to demonstrate proper tolerance  against noise and errors in scaling up the number of qubits \cite{{Gambetta17},{Martinis18}}. 

As the number of qubits grows it becomes more difficult to control quantum states with high precision. Achieving high fidelity  control not only requires improvements in the device fabrication \cite{Martinins16}, but also calls for  rectifying  theoretical estimation \cite{{Preskill},{theory}}. Currently some well known models in quantum optics are used to describe superconducting circuits, such as multilevel Jaynes-Cumming model and its generalizations \cite{JC}. Superconducting circuit theory should gain ability in going beyond the regimes admissible by current models and  become safely extendible to massive qubit lattices \cite{{Wallraff11},{Nataf}}.

\begin{figure}[t]
\vspace{-0.5cm}
		\includegraphics[width=0.77\linewidth]{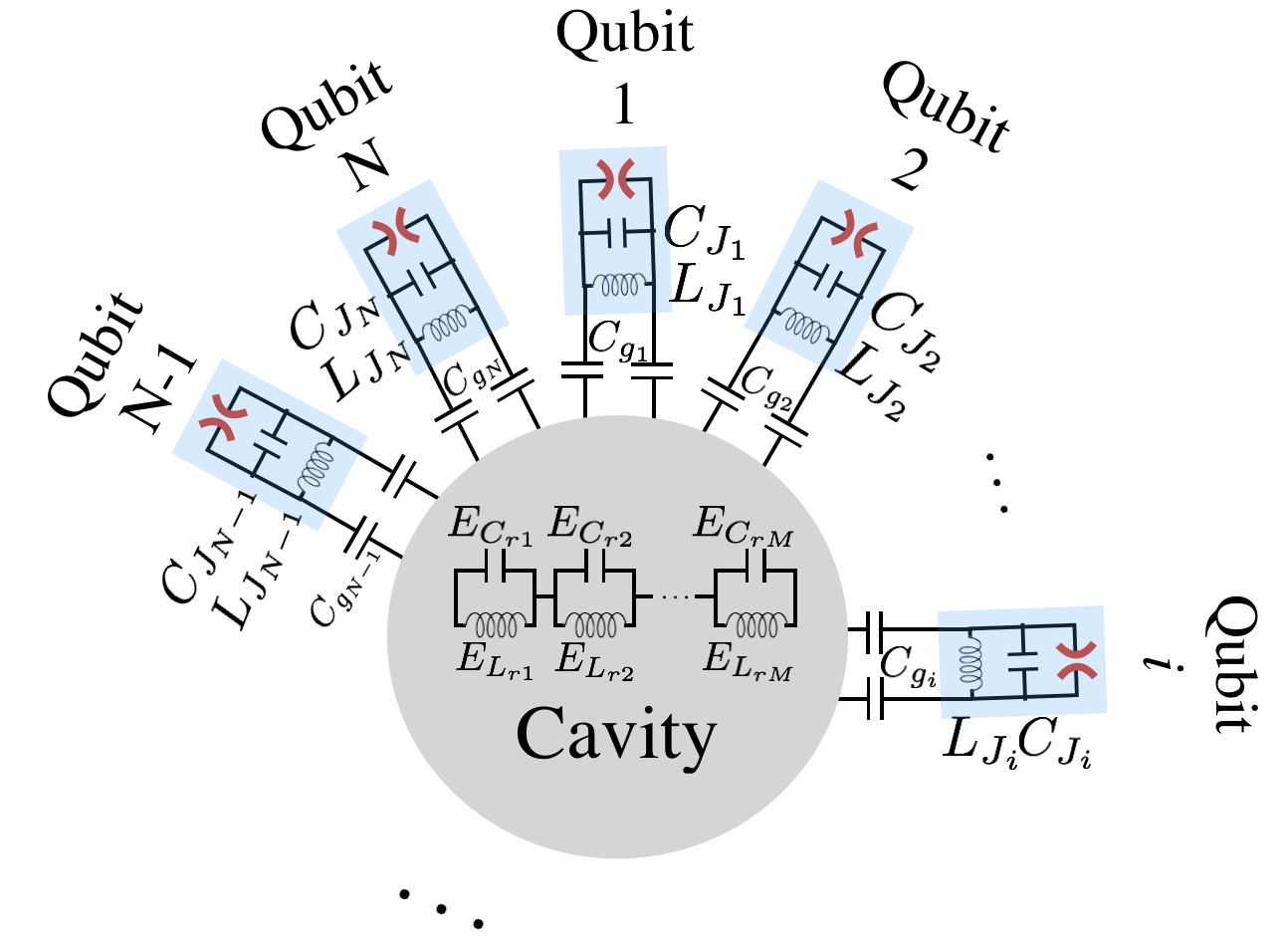}
		\vspace{-0.4cm} \caption{$N$ weakly anharmonic qubits (blue boxes)  coupled to a multimode cavity.  Curly (red) crosses indicate anharmonicity and the  coupled LC circuits make the harmonic sector.}
		\label{fig schem}
	\end{figure}

Recently a formalism has  been introduced to consistently quantize weakly anharmonic JJs coupled to cavity modes \cite{Nigg12}. These circuits will be of our interest in here too. The so called  \emph{black box  quantization} divides the circuits into two sectors: the harmonic sector and the anharmonic one.   The latter has been indicated in Fig.(\ref{fig schem})  by  (red) curly crosses of JJs, and the remaining LC circuits make the harmonic sector.  In the lack of nonlinearity   JJs and cavity modes can be treated on equal footing, therefore they are lumped into effective impedances seen by the anharmonic sector. Identifying these effective impedances is, however, of the central importance in this quantization, for which Nigg et.al. in Ref. \cite{Nigg12} employ a pole-decomposition technique, which simplifies the harmonic sector into a Foster-equivalent LC circuits \cite{Foster}. Numerical  evaluation of the impedances takes place  in iterative feedbacks between experiment and theory.  The circuit is then quantized to properly include anharmonicy. So far this formalism has successfully  described theoretical issues, such as in  cut-off free coupling to a multimode cavity \cite{tureci17}, inductively shunted transmon \cite{Richer17}, and dispersive interaction rates \cite{Firat17}.

Motivated by black box quantization, here we determine a normal mode basis, from which one can obtain all of the seemingly-independent effective impedances theoretically.  We  study the two examples of one  or two of the weakly anharmonic transmons coupled to resonator. We determine  physical parameters in closed  form formulas, which will remain valid at arbitrary coupling strengths and bare frequencies. The large domain of validity can be otherwise reproduced by a combination of various models each of which only valid within a limited domain. Therefore black box quantization is natural and indispensable quantizastion method for such circuits. 
%

\emph{A transmon coupled to a resonator case. --} In a Cooper pair box (CPB)  made of a transmon coupled to single-mode resonator, canonical variables are charges and phases, i.e. $({q}_i, {\phi}_i)$  with $i$ being $t$ (transmon) or $r$ (resonator) \cite{Clerk}.  Interaction takes place by gate capacitor $C_g$ that capacitively couples transmon to the center conductor of resonator. With the transmon charge being exposed to the resonator voltage $V_r=q_r/C_r$, the dipole interaction is $H_{int}=\beta V_r q_t$, with  $C_{r/t}$  being resonator/transmon capacitances and $\beta\equiv C_g/C_t$. Keeping $\beta \ll 1 $  is essential in order to keep qubit coherence time sufficiently long \cite{Manucharyan}.  The Hamiltonian is   
\begin{eqnarray} \nonumber
H=H_{\textup{har}}+H_{\textup{anhar}}, \ \ \ \ H_{\textup{anhar}}=- \frac{E_C }{3 Z_{t}^2 \hbar^2}  \phi_t^4 && \\ 
     H_{\textup{har}}=\sum_{i=r,t} \frac{q_i^2}{2C_i} + \frac{\phi_i^2}{2L_i}  +H_{int} &&   \label{eq. Hexm1}
\end{eqnarray}
with  characteristic linear impedances $Z_{i}=\sqrt{L_i/C_i}$ and harmonic  frequencies $\omega_{i}=1/\sqrt{L_i C_i}$, $E_C$ being total capacitive energy of transmon (including JJ and shunt capacitances as well as any capacitive coupling  between transmon and voltage sources),  and $\hbar$ the reduced Planck constant.   We define  canonical variables $({Q}_i, {X}_i)\equiv ({q}_i {L_i}^{1/2},{\phi}_i {L_i}^{-1/2})$. In this  basis the charge and phase vectors are  $\mathbf{Q}=(Q_t,Q_r)$ and  $\mathbf{X}=(X_t,X_r)$, respectively and the harmonic part of Eq. (\ref{eq. Hexm1})  is transformed into  
\begin{equation}
H_{\textup{har}}=\frac{1}{2}\mathbf{Q}^\textup{T} \mathbf{M Q}+\frac{1}{2}\mathbf{X}^\textup{T}\mathbf{  X}; \  \mathbf{M}=\left[\begin{array}{ccccc}
\omega_{t}^{2} &  g \sqrt{4\omega_{t}\omega_{r}}\\
g \sqrt{4\omega_{t}\omega_{r}} & \omega_{r}^{2}  \label{eq:M}
\end{array}\right],
\end{equation}
with $g \equiv \beta \omega_r \sqrt{Z_r/4Z_t}$. 

This Hamiltonian can be diagonalized by unitarily transforming $\mathbf{Q}$ and $\mathbf{X}$ into new canonical variable $\mathcal{Q}$ and $\mathcal{X}$, i.e.  ${Q}_i=\sum_j S_{ij} \mathcal{Q}_j$ and ${X}_i=\sum_j T_{ij} \mathcal{X}_j$.  Given that the variables in  the new and the old frames must satisfy the Poisson brackets of canonical coordinates, i.e. $\{\mathcal{Q}_i,\mathcal{X}_j\}=\{Q_i,X_j\}= \delta_{ij}$,  one can  find  that $T=S$   (see Supplementary Material  \ref{sec.Unitary_SM}).  The only term in Eq. (\ref{eq:M}) that should be diagonalized  is  $\mathbf{Q}^\textup{T}\mathbf{ M Q}$, which in the new basis must look like  $\mathcal{Q}^\textup{T} \Omega \mathcal{Q}$  with $\Omega$ being a diagonal matrix,  $\Omega_{tt}=\bar{\omega}_t^2$, $\Omega_{rr}=\bar{\omega}_r^2$ and zero otherwise.  This brings up  the important conclusion that the unitary transformation  $\mathbf{S}$ is indeed the matrix made of columns of normalized eigenvectors of $\mathbf{M}$.  

The transformation of harmonic sector to a basis with uncoupled harmonic oscillators determines the following frequencies for the linear transmon and resonator: $\bar{\omega}_t=K_-^2$ and $\bar{\omega}_r=K_+^2$, respectively, with $K_\pm \equiv 2^{-\frac{1}{4}} \left( \omega_t^2+\omega_r^2\pm \Delta \Sigma s^{-1}\right)^{\frac{1}{4}}$ and $s\equiv ({1+{16 \left({g}/{\Delta}\right)^2 \omega_r \omega_t}/{\Sigma^2} })^{-1/2}$,  $\Sigma\equiv \omega_r+\omega_t$ and $\Delta\equiv \omega_r-\omega_t$.  As discussed above, the unitary transformation $\mathbf{S}$  is the columns of normalized eigenvectors, which are $(\pm \sqrt{({1\mp s})/2}, \sqrt{({1\pm s})/2})^T$ associate to the eigenvalues $K_{\pm}^2$.  (These solutions  can be confirmed for analogue quantum circuits using Bogoliubov transformations \cite{BV}, for details see Supplementary Material \ref{sec.Bogoliuboc_SM}.)

After diagonalizing the harmonic sector classically and obtaining the normal mode basis, the anharmonic sector can  be transformed into the same basis.  Given that $\mathbf{S}$ matrix transforms phases between the two basis, i.e. $X_t= -\sqrt{(1+s)/2}\mathcal{X}_t+\sqrt{(1-s)/2}\mathcal{X}_r$, and that anharmonicity  depends on $X_t^4$ one can see that all sort of interactions is possible, e.g. $C_{m} \mathcal{X}_r^m \mathcal{X}_t^{4-m}$ with  coupling strengths $C_{m}(s)$ and $m=0,1,2,3,4$.   

The quantization of $Q_j,X_j$ with $j=t,r$ can take place by redefining them in terms of ladder operators: $\hat{Q}_j= \sqrt{\hbar /2 \omega_j} (\oad_j + \oa_j) $ and $\hat{X}_j=i\sqrt{\hbar \omega_j/2 } (\oad_j - \oa_j)$, with $\hat{a}_j=\sum_{n_j} \sqrt{n_j+1} \ket{n_j}\bra{n_j+1}$ making transitions between the energy eigenbasis $|n_j \rangle$ \cite{Clerk}. Rewriting the new phase variables in terms of the ladder operators $\hat{\alpha}_k$ defined in the normal mode basis, one can obtain the following Bogoliubov transformation between old and new bases: $\oad_t-\oa_t=U_{tt}(\oald_t - \oal_t) +U_{tr}(\oald_r - \oal_r)$, with $U_{tt}=-[{ (1+s){\bar{\omega}_t}/{2}{\omega_t}}]^{\frac{1}{2}}$ and $U_{tr}= [{ {(1-s)} {\bar{\omega}_r}/2{{\omega}_t}}]^{\frac{1}{2}}$. The Hamiltonian of anharmonic sector in Eq. (\ref{eq. Hexm1}) can be represented in quantum theory by $H_{\textup{anhar.}}=-(\delta/12) (\oad_t-\oa_t)^4$ with  the anharmonic coefficient  $\delta\equiv  E_C$.  In the new basis this will be transformed to $-\frac{\delta}{12} (U_{tt}(\oald_t - \oal_t) +U_{tr}(\oald_r - \oal_r))^4$, which can be used to define self-Kerr coefficients  ${\chi}_i$  using the following general form: $-({1}/{12}) ( {\chi}_t^{{1}/{4}} (\oald_t-\oal_t) +{\chi}_r^{{1}/{4}} (\oald_r-\oal_r)  )^4$ \cite{Boissonneault}.   Notice that anharmonicity is not diagonal in the new basis, however  we can simplify the anharmonic Hamiltonian by ignoring irrelevant terms to the first order anharmonicity and  applying secular approximation. This reformulates the  Hamiltonian to  
\begin{eqnarray} \nonumber \label{eq. H_1t_1r}
H&=&\sum_{i=t,r}\bar{\omega}_{i} \oal_i^{\dagger} \oal_i-\frac{{\chi}_{i}}{2}   \left[ \left( \oal_i^{\dagger} \oal_i \right)^{2} + \oal_i^{\dagger} \oal_i +\frac{1}{2}\right]  \\ &&   - 2 {\chi}_{rt} \left( \oal_t^{\dagger} \oal_t  +\frac{1}{2}\right) \left( \oal_r^{\dagger}\oal_r  +\frac{1}{2}\right), 
\end{eqnarray}
in which the cross-Kerr  is defined  ${\chi}_{rt}\equiv \sqrt{{\chi}_{r}{\chi}_{t}}$. One can see in Eq. (\ref{eq. H_1t_1r}) that transforming JJ nonlinearity into the normal mode basis introduces weak interaction between transmon and resonator with strength being the cross-Kerr coefficient and is linearly proportional to anharmonicity.

Let us define \emph{dressed frequency} $\tilde{\omega}_i$ to be the coefficient of $\hat{\alpha}_i^\dagger \hat{\alpha}_i$. By summing similar terms one can obtain the following closed form formula for the transmon and resonator dressed frequencies: 
\begin{eqnarray}  
&& \tilde{\omega}_{t} =  K_-^2 -\frac{{\chi}_t}{2}-{\chi}_{rt}, \ \ \ \ \tilde{\omega}_{r}  =  K_+^2 -\frac{{\chi}_r}{2}-{\chi}_{rt}  \label{eq. freqs}
\\ &&  {\chi}_{t} =  \delta{\left(1+s\right)^{2}}K_{-}^4/{4\omega_t^{2}}, \ \  {\chi}_{r}  =  \delta {\left(1-s\right)^{2}}{K_{+}^{4}}/{4 \omega_t^2}\ \ \  \ \   \label{eq. kerr}
\end{eqnarray}
which is valid  at arbitrary value of $g/\Delta$.  This leads to the energy levels $E_{n_t n_r} \simeq \sum_{i=t,r}\tilde{\omega}_i n_i-(\chi_i^2/2)n_i^2-2 \chi_{rt} n_t n_r$. 

In dispersive regime, in which detuning  frequency between transmon and resonator is much stronger than the coupling strength, i.e. $g/\Delta \ll 1$,  Eqs.  (\ref{eq. freqs}) and (\ref{eq. kerr}) can be expanded in all orders of $g/\Delta$. This will result in the self-Kerr  coefficients  ${\chi}_{t}  =  \delta [1- 4  ({g}/{\Delta})^{2} \omega_r (\omega_r^2+\omega_q^2) /\omega_q \Sigma^2]$ and   ${\chi}_{r} =   16 \delta \left({g}/{\Delta}\right)^{4} \omega_r^4 / \Sigma^4 $. The transmon and resonator  dressed frequencies will become  $\tilde{\omega}_{t} \approx \omega_{t} - 2 \omega_r  {g}/{\Delta}\Sigma$, $ \tilde{\omega}_r  \approx \omega_{r}+ 2 {g^2}\omega_t/{\Delta}\Sigma  $.   These expression are in agreement with  the non-Rotatating Wave Approximation (non-RWA)  results recently taken in the second order perturbation theory \cite{Gely}.  In circuits with $g\ll \Delta \ll \Sigma$, RWA can  simplify the Kerr coefficients. By  applying the approximation, one gets ${\chi}_{t}^{\textup{RWA}}  \approx  \delta [1- 2  ({g}/{\Delta})^{2} ]$,    ${\chi}_{r}^{\textup{RWA}} \approx   \delta \left({g}/{\Delta}\right)^{4} $,   ${\chi}_{rt}^{\textup{RWA}}\approx \delta \left( {g}/{\Delta}\right)^2   $, and  dressed frequencies: $\tilde{\omega}_{t}^{\textup{RWA}} \approx \omega_{t}-{\delta}/{2} -   {g^2}/{\Delta} - \delta g^2/ \Delta^2 $,  $\tilde{\omega}_r^{\textup{RWA}}  \approx \omega_{r}+  {g^2}/{\Delta}  - \delta g^2/\Delta^2 $, which confirm the original perturbative Lamb and AC-Stark shifts  reported by J. Koch, et.al. in Ref. \cite{Koch07} and observed in \cite{Fragner}.

\begin{figure}[t]
\begin{center}
 \includegraphics[width=0.77\linewidth]{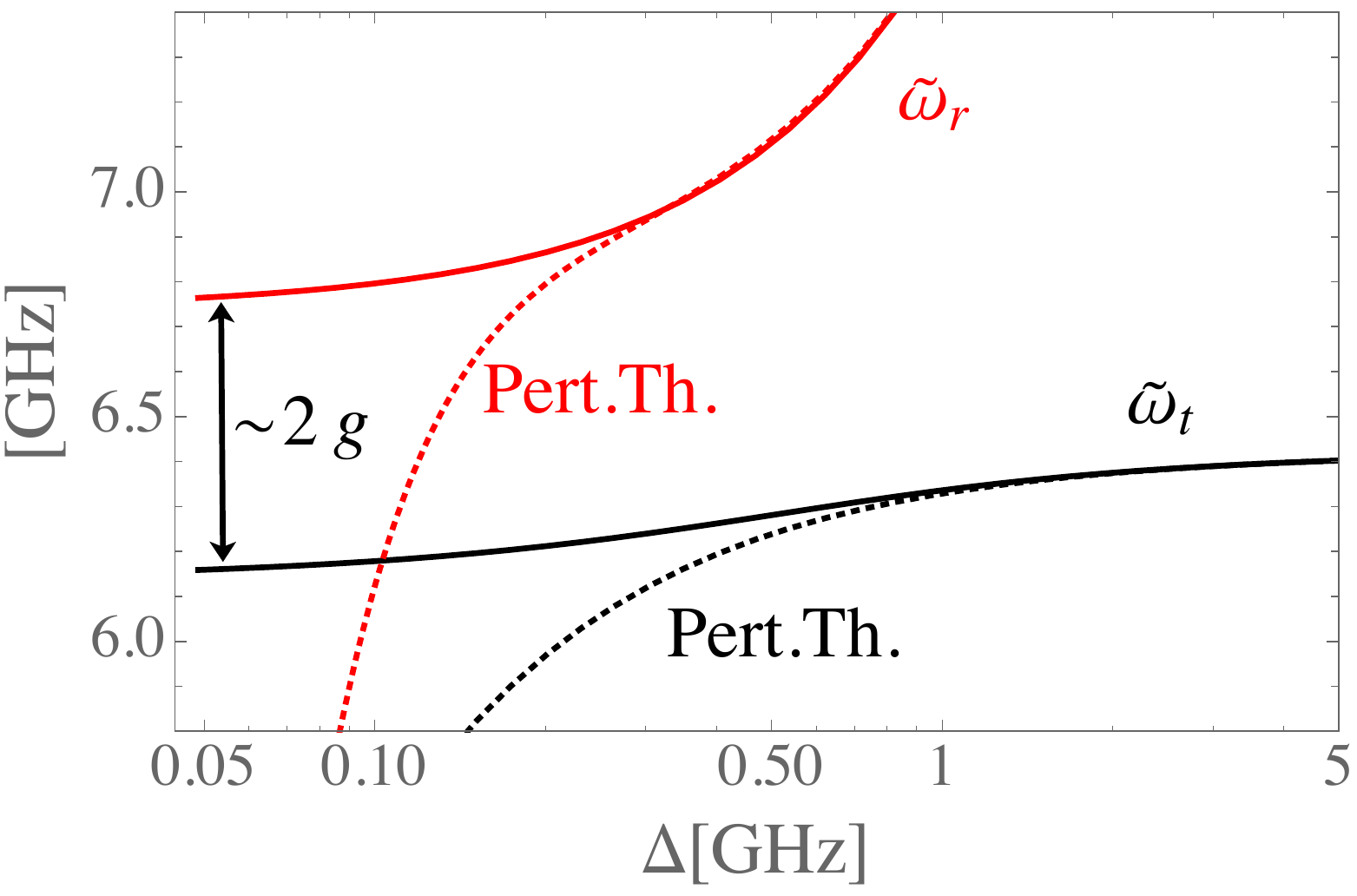} \\  
  \includegraphics[width=0.77\linewidth]{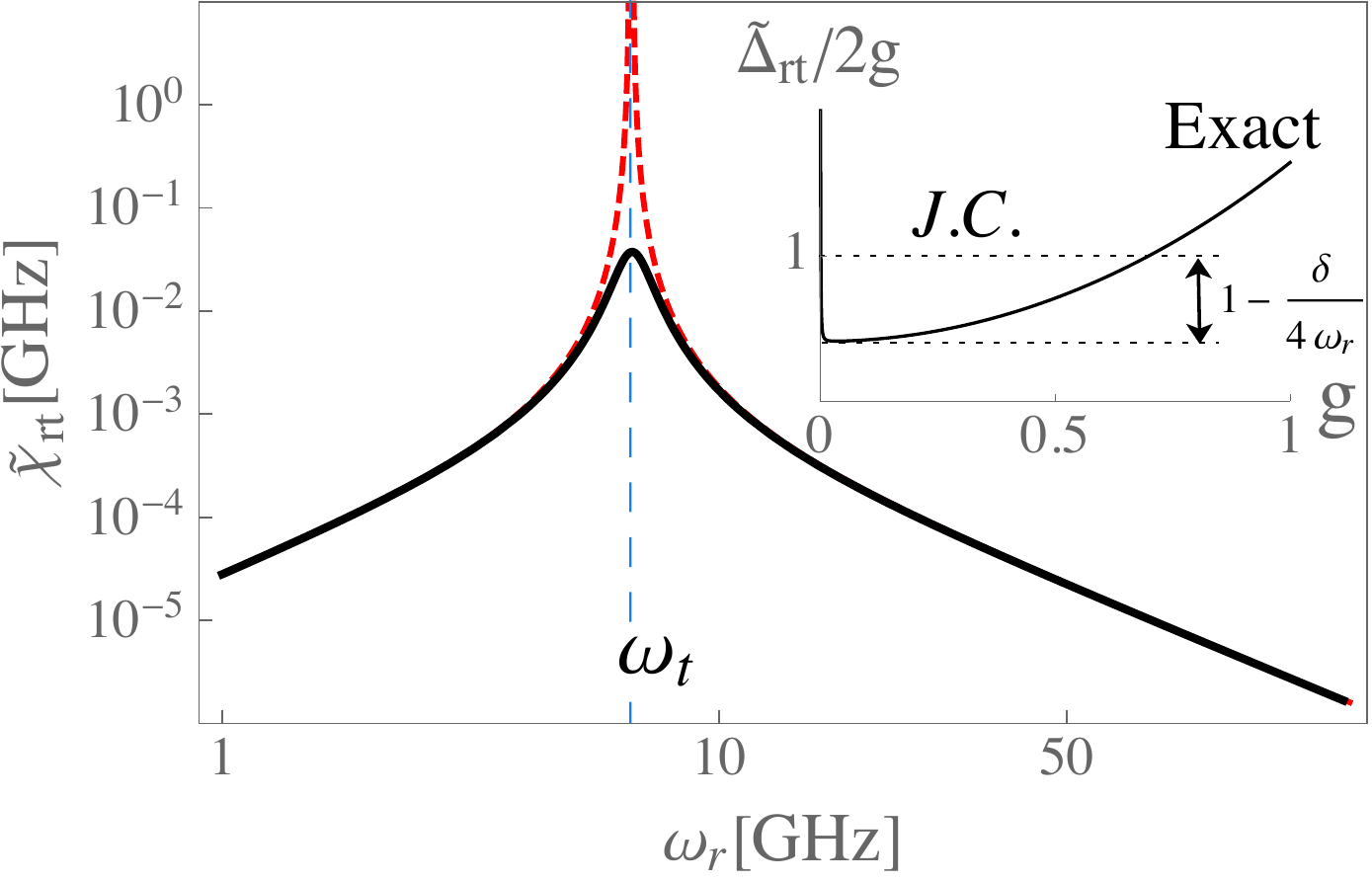} 
\caption{Exact (solid) and perturbative (dashed) results for (a) dressed frequencies  (b)  cross Kerr in circuit with $\omega_t=6.5$GHz, $\omega_r=\Delta+\omega_t$, $g=0.3$GHz, and $\delta=0.15$GHz. (b Inset) Dressed frequency detuning  at the resonant point $\Delta=0$ from Jayns-Cumming model (JC) and exact model (solid).}
\label{fig 1}
\end{center}
\end{figure}

Fig(\ref{fig 1}a) shows dressed resonator and transmon frequencies with respect to frequency detuning $\Delta$ and compares exact (solid) frequencies  of Eq. (\ref{eq. freqs}) with perturbative (dotted) frequencies. In these circuit parameters the mismatch between RWA and non-RWA results are negligible in the logarithmic scales.    Far away from the dispersive regime and near resonance $\Delta\approx 0$, where perturbative theory diverges, multilevel Jayns-Cumming model \cite{JC} seems to be a reliable model \cite{Haroche}. At zero detuning limit the  degeneracy of the two harmonic frequencies are lifted by $2g$ due to atom-photon coupling \cite{Blais04}.  Interestingly  our results of  Eq. (\ref{eq. freqs}) naturally gives rise to the frequency detuning $\tilde{\Delta}_{rt} = 2g(1 - \delta /4\omega_r) + O(g^2)$ with $\tilde{\Delta}_{rt}\equiv \tilde{\omega}_r-\tilde{\omega}_t$.  This not only meets our expectations, but also provides correction in the degeneracy lifting value due to the presence of anharmonicity. In Fig. (\ref{fig 1}b inset) we show $\tilde{\Delta}_{rt}/2g$, which is 1 in Jaynes-Cumming model; however Eq. (\ref{eq. freqs}) shows that in the small $g\ll\omega_r$ limit it is $1 - \delta /4\omega_r$ and in large $g$ it is nonlinear in $g$.  

Fig. (\ref{fig 1}b) plots exact cross Kerr coefficient defiened below Eq. (\ref{eq. kerr}) and compares it with perturbative (dashed) results. At the resonant point, i.e. $\omega_r \approx \omega_t$, perturbation theory diverges as expected, however our results shows that the divergence is not physical . Interestingly at resonant point cross Kerr coefficient is the universal value $\delta/4+O(g^2)$ no matter how much are transmon and resonator bare frequencies.


\emph{$N$-atoms coupled to $M$ resonator. --}  As shown in Fig. (\ref{fig schem}) $N$ weakly-anharmonic transmon modes individually interact  with $M$ cavity harmonic modes. Such a circuit in applicable for measuring entanglement scaling, sensing, quantum  computation, etc.  

There are $2(N+M)$   canonical variables for charge and phase degrees of freedom: $\mathbf{Q}=\left(Q_{1},\cdots,Q_{N+M}\right)^{T}$ and  $\mathbf{X}=\left(X_{1},\cdots,X_{N+M}\right)^{T}$. The generalized harmonic Hamiltonian is $H_{\textup{har.}}=\frac{1}{2}\sum_{i=1}^{N+M} \omega_i^2 Q_{i}^{2}+\frac{1}{2} X_{i}^{2}+\sum_{i=1}^{N} \sum_{j=N+1}^{N+M} g_{ij}\sqrt{4\omega_{i}\omega_{j}}Q_{i}Q_{j}$, which can be reformulated into $H_{\textup{har.}}=\frac{1}{2}\mathbf{Q}^\textup{T} \mathbf{M Q}+\frac{1}{2}\mathbf{X}^\textup{T}\mathbf{  X}$, with generalized $(N+M)\times (N+M)$ matrix $\mathbf{M}$. Defining indices $t=\{ 1,2, \cdots,N\}$ for transmon subspace and $r=\{N+1,\cdots,M\}$ for resonator subspace, there are the following nonzero arrays at $M_{rr}=\omega_r^2$,  $M_{tt}=\omega_a^2$, $M_{t r}=M_{r t}=g_{t}\sqrt{4 \omega_t \omega_r}$, and zero otherwise.   

All we need is to diagonalize this harmonic sector. For this aim we define a new frame in which the Hamiltonian becomes $\frac{1}{2}\sum_{i}\bar{\omega}_i\mathcal{Q}_{i}^{2}+\mathcal{X}_{i}^{2}$ using the following unitary transformations ${Q}_{i}=\sum_{j}S_{ij}{\mathcal{Q}}_{j}$ and ${X}_{i}=\sum_{j}T_{ij}{\mathcal{X}}_{j}$. As  discussed in previous section  the unitary transformation of these canonical variables satisfy $T=S$, noticing that $S$ is the matrix of normalized eigenvectors of $\mathbf{M}$ with columns being eigenvectors (see Supplementary Material  \ref{sec.Unitary_SM}). 

Using the definition of ladder operators similar to what proposed above Eq. (\ref{eq. H_1t_1r})  one can obtain Bogoliubov transformation between ladder operators in non-diagonal and diagonal harmonic bases
\begin{eqnarray}
\hat{a}^\dagger_{i}-\hat{a}_i & = & \sum_{j=1}^{N+M}  U_{ij}\left( \hat{\alpha}^\dagger_{j}- \hat{\alpha}_j \right), \ \ \ U_{ij}\equiv \sqrt{\frac{\bar{\omega}_j}{\omega_i}} S_{ij}  \label{eq. Bog}  
\end{eqnarray}
The anharmonic Hamiltonian is $H_{\textup{nonhar.}}=\sum_{i=1}^{N}(\delta_i/12)(\hat{a}_i-\hat{a}_i^\dagger)^4$ which should be transformed to the new basis using Eq. (\ref{eq. Bog}). For explicit evaluation of the anharmonic terms in the transformed basis see Supplementary Material \ref{sec. anharmonicity_SM}. In the following we consider another example of two interacting transmons coupled to a resonator.


\emph{Example: Two transmons coupled to a resonator.--} This circuit is used in recent experiments specially for making 2 qubit gates, which is a challenging research in quantum computation \cite{{Blais07},{Majer07}}.   Let us denote transmons and resonator bare frequencies  $\omega_i$ with $i=1,2,3$, respectively; however sometimes we use index $r$ instead of $3$ to emphasize on the resonator.   Interactions take place with the strength couplings  $g_{1}, g_{2}$ between the transmons and the resonator.  The $3\times 3$ matrix $\mathbf{M}$ is
\begin{equation}
\mathbf{M}=\left[\begin{array}{ccc}
\omega_{1}^{2} & 0 & g_{1}\sqrt{4\omega_{1}\omega_{3}}\\
0 & \omega_{2}^{2} & g_{2}\sqrt{4\omega_{2}\omega_{3}}\\
g_{1}\sqrt{4\omega_{1}\omega_{3}} & g_{2}\sqrt{4\omega_{2}\omega_{3}} & \omega_{3}^{2}
\end{array}\right]\label{eq:M3dar3}
\end{equation}

Finding the eigenvalues of this matrix is cumbersome; however,  within a certain domain of parameters, which is wide enough to cover the circuits of interest (see below Eq. (\ref{eq:eignfreq})), we can find eigenvalues analytically. The cubic equation that determines eigenvalues of Eq. (\ref{eq:M3dar3}) is $\lambda^{3}+b\lambda^{2}+c\lambda+d=0$ with $\lambda$ being eigenvalues, 
$b\equiv -\sum_{i=1,2,3}\omega_{i}^{2}$,
$c\equiv  \omega_{1}^{2}\omega_{2}^{2}+\omega_{1}^{2}\omega_{3}^{2}+\omega_{2}^{2}\omega_{3}^{2}-\sum_{i=1,2} 4g_{i}^{2}\omega_{i}\omega_{3}$,
and $d\equiv  4g_{2}^{2}\omega_{1}^{2}\omega_{2}\omega_{3}+4g_{1}^{2}\omega_{1}\omega_{2}^{2}\omega_{3}-\omega_{1}^{2}\omega_{2}^{2}\omega_{3}^{2}$. Let us define new variables $\theta\equiv \lambda+b/3$ which help eliminate quadratic term. The new equation looks like $\theta^{3}-f\theta+h=0$ with $f\equiv  b^{2}/3-c$ and $h\equiv  \left(2b^{3}-9bc+27d\right)/27$. We find the eigenvalues $\lambda$ of the matrix $\mathbf{M}$ in Eq. (\ref{eq:M3dar3})  and given that $\bar{\omega}_{k}=\sqrt{\lambda_{k}}$, 
\begin{equation}
\bar{\omega}_{k}^2 =  {2\sqrt{\frac{f}{3}}\cos \frac{ \textup{cos}^{-1}\left(-\frac{h}{2}\left(\frac{3}{f}\right)^{\frac{3}{2}}\right)-2\pi(k-1)}{3}-\frac{b}{3}}  \label{eq:eignfreq}
\end{equation}
with  $k=1,2,3$. Notice that proper relabelling of indices might be required to identify what frequencies are those of transmons and of the resonator. The argument of  $\cos^{-1}$ function must stay between 1 and -1, which enforces the following condition ${h^{2}}/{4}-{f^{3}}/{27}<0$ to be satisfied (see Supplementary Material \ref{sec.limits_SM} for further details).  Given that frequencies are at least  an order of magnitude larger that interaction, the condition is trivially satisfied and from Eq. (\ref{eq:eignfreq}) three real-values dressed frequencies can be expected.

The anharmonic sector should be transformed into this normal mode basis. The transformation matrix is the columns of normalized eigenvectors of Eq. (\ref{eq:M3dar3}), whose explicit form can be found in Supplementary Material \ref{sec. trans 2transmons_SM}.  In the new basis we keep only secular terms and terms with  preserve excitation number. In the first order of $\delta$ the circuit Hamiltonian becomes 
\begin{eqnarray} \nonumber
H&=&\sum_{i=1,2,3} \tilde{\omega}_i \hat{\alpha}_i^\dagger \hat{\alpha}_i       \\ \nonumber &&
  - \sum_{i=1,2,3}\Bigg\{ \frac{\chi_{i}}{2}\left(\hat{\alpha}_{i}^{\dagger}\hat{\alpha}_{i}\right)^{2} +  2 \sum_{k > i} \chi_{ik}\left(\hat{\alpha}_{i}^{\dagger}\hat{\alpha}_{i}\right)\left(\hat{\alpha}_{k}^{\dagger}\hat{\alpha}_{k}\right)  \Bigg.  \\ \nonumber &&
 \ \ \ +   \sum_{k > i} \left( \mathcal{J}_{ik}+ \sum_{l\neq i,k} \mathcal{L}_{ikl} \hat{\alpha}_{l}^{\dagger}\hat{\alpha}_{l}  \right) \left(\hat{\alpha_{i}}\hat{\alpha}_{k}^{\dagger}+\hat{\alpha}_{i}^{\dagger}\hat{\alpha}_{k}\right)
   \\  && \Bigg.  \ \ \ + \sum_{k\neq i} \mathcal{K}_{ik}  \left[\left(\hat{\alpha}_{i}^{\dagger}\hat{\alpha}_{i}\right)\hat{\alpha}_{i}\hat{\alpha}_{k}^{\dagger}+\hat{\alpha}_{i}^{\dagger}\hat{\alpha}_{k}\left(\hat{\alpha}_{i}^{\dagger}\hat{\alpha}_{i}\right)\right] \Bigg\}
  \label{eq. 2trans}
 \end{eqnarray}
with self-Kerr $\chi_{i}=\sum_{j=1,2} \delta_j U_{ji}^{4}$ and  cross-Kerr $\chi_{ik}=\sum_{j=1,2} \delta_j U_{ji}^{2} U_{jk}^{2}$ and $U_{ij}$ being define in Eq. (\ref{eq. Bog}).  One can easily evaluate all  self and cross Kerr cofactors and check  that  in general there is no simple relation between cross Kerr  and self-Kerr coefficients.  The dressed frequency of   transmons and  resonator are 
\begin{equation}
\tilde{\omega}_{i} = \bar{\omega}_{i} -\frac{{\chi}_i}{2}-\sum_{j  (\neq i)}{\chi}_{ij}. \label{eq. dress_w_2tr}
\end{equation}

\begin{figure}[t]
\begin{center}
 \includegraphics[width=0.77\linewidth]{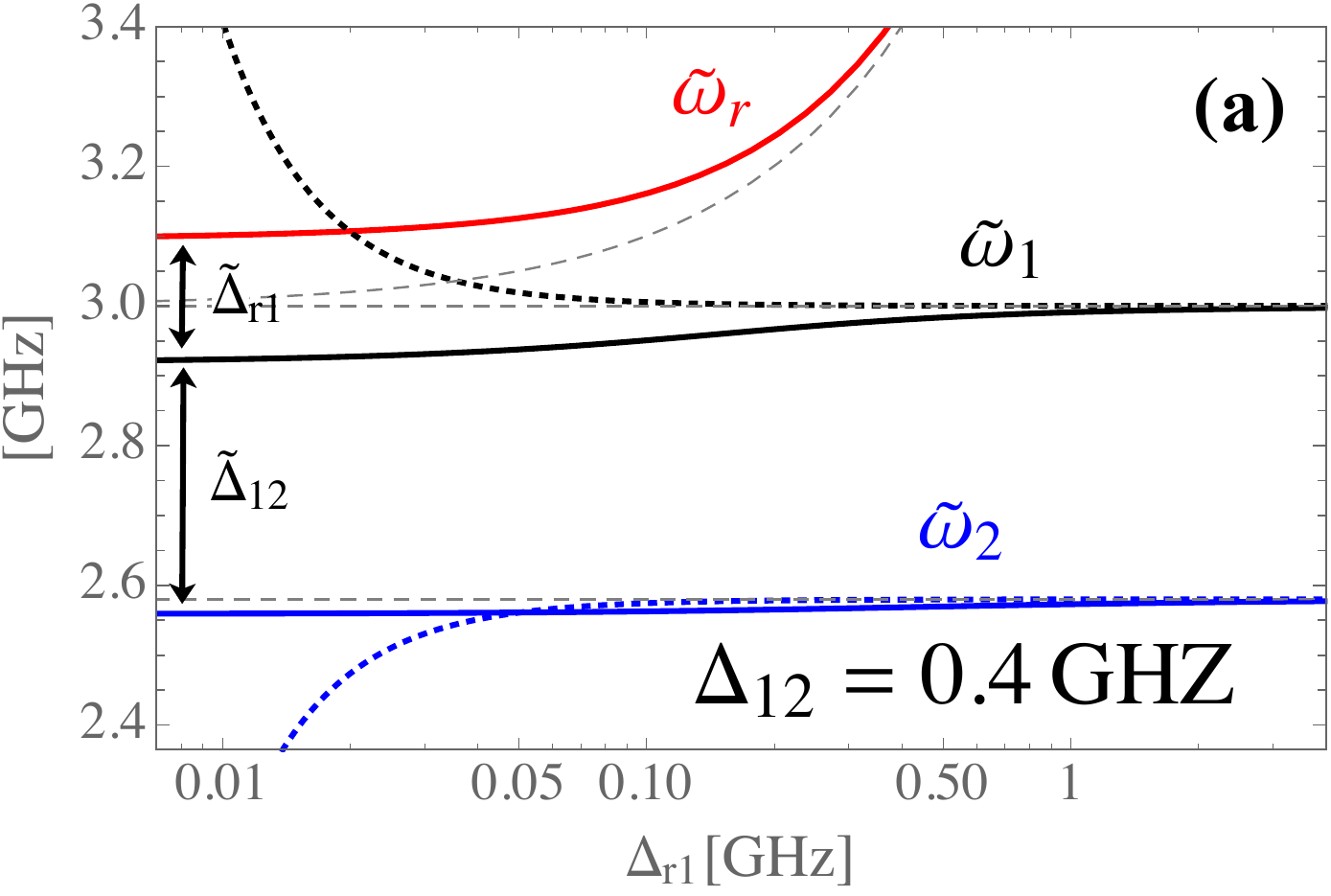} \\ 
  \includegraphics[width=0.77\linewidth]{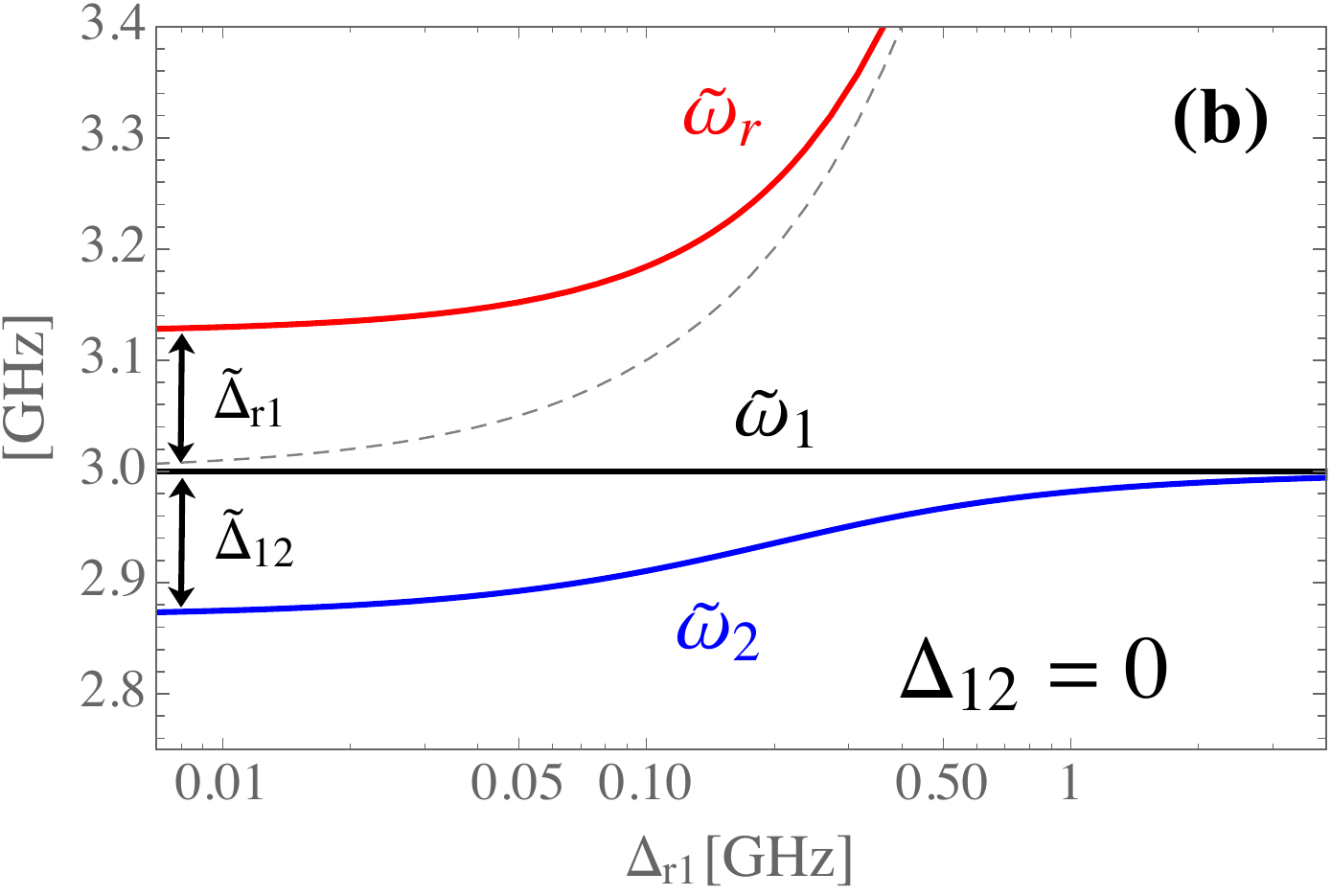} 
\caption{Perturbative (dotted) and exact (solid)  Dressed frequencies in circuit with bare frequencies (dashed) of 2 transmons $\omega_1=3$GHz and $\omega_2=\alpha \omega_1$, and a resonator $\omega_r=\omega_1+\Delta_{r1}$, couplings $g_{1/2}=0.1$GHz, and anharmonicity $\delta_{1/2}=0.1$GHz. (a) $\alpha=0.86$. (b) Resonant transmons $\alpha=1$. }
\label{fig 3}
\end{center}
\end{figure}

Let us make further elaboration on the second two lines in Eq. (\ref{eq. 2trans}). The term with the coupling  $\mathcal{J}$ indicating direct anharmonic interaction between two oscillators.  $\mathcal{K}$ and $ \mathcal{L}$ couplings both are multiplied by $(\hat{\alpha}^\dagger_i \hat{\alpha}_i)$, therefore the coupling strengths  are in fact $ n_l \mathcal{L}_{ikl} $ and $n_i \mathcal{K}_{ik} $ with $n$ being  integer quantum numbers; thus they introduce the contribution of higher excitations in anharmonic interaction. Explicit definitions of these couplings can be found in Supplementary Material \ref{sec_Additional}, which shows they are linearly proportional to the anharmonicity $\delta$. Block-diagonalization of this Hamiltonian ---using for example Schrieffer-Wolff \cite{SW} (see also Supplementary Material \ref{sec. SW_SM} for a brief overview)---  indicates that  computational states carry the contribution of these interactions in the second order of $\delta$. Therefore all these couplings  can be safely disregarded for the evaluation of dressed frequencies in the leading order of $\delta$.  
The energy eigenvalue in the first order of $\delta$ is $E_{n_1 n_2 n_3} \simeq \sum_{i=1,2,3} n_i \tilde{\omega}_{i}-(\chi_i/2) n_i^2-2\sum_{k> i}\chi_{ik} n_i n_k $.

Consider a circuit with transmon bare frequencies $\omega_1$ and $\omega_2=\alpha \omega_1$, and the resonator frequency  $\omega_r=\omega_1+\Delta_{r1} $. Fig. (\ref{fig 3}a) shows dressed frequencies for $\omega_{1}=3$GHz and $\alpha=0.86$ obtained from perturbation theory (dotted) and Eq. (\ref{eq:eignfreq})  (solid).  For the perturbative results  we used the approach explained in \cite{Gambetta13} and references therein. In the dispersive regime both approaches give rise to almost similar results. However, near the resonance $\Delta_{r1}\approx 0$ black box quantization results in  finite dressed frequencies, while  perturbation theory diverges. For $\alpha<1$  by expanding Eq. (\ref{eq:eignfreq}) in the limit of $g/ \omega_1 < (1-\alpha^2)/3\sqrt{6}$  dressed frequencies can be obtained in different orders of $g$. In absence of anharmonicity one can find the following detuning of dressed frequencies:  $\tilde{\Delta}_{r1}\approx 2g+O(g^3)$ and $\tilde{\Delta}_{12} \approx \Delta_{12}-g + [(1+\alpha)^{-1}+(1+\alpha)/2]g^2/\Delta_{12}+O(g^3)$.

Resonant transmons are studied in Fig. (\ref{fig 3}b), where there is no  difference between  transmon bare frequencies  $\alpha=1$, therefore perturbation theory is not applicable. However Eq. (\ref{eq:eignfreq}) determines the following finite values for dressed frequencies: $\omega$, $[({\omega^2+\omega_r^2 \pm \Delta \Sigma r^{-1}})/2]^{1/2}$ with $r^{-2}\equiv 1+32 g^2 \omega \omega_r/\Delta_{r1}^2 \Sigma_{r1}^2$. At the extreme resonance with both transmons and the bus resonator having the same bare frequencies, i.e. $\Delta_{12}=\Delta_{r1}=0$, detuning  between the dressed frequencies turn out to be $\tilde{\Delta}_{r1} \approx \tilde{\Delta}_{12}\approx \sqrt{2} g$.  In supplementary Material \ref{sec. resonance_SM} we derive these results using a direct diagonalization.

Before conclusion let us  further comment on the black box quantization. Originally  \cite{Nigg12} introduced the effective impedances by which one can find the normal mode basis of the harmonic circuit.  However, theoretically these impedances have been  left undetermined and evaluating them requires feedback from experiment in the model. Here, instead we introduced a unitary transformation that diagonalizes linear LC circuits. For a transmon coupled to a resonator, one can find that  $Z_q^{\textup{eff}}=\tilde{\omega}_q {(1+s)Z_{t}}/ {2\omega_q}$ and  $Z_r^{\textup{eff}}=\tilde{\omega}_r {(1-s)Z_{t}}/{2\omega_q}$. The ratio of the two impedances is $Z_r^{\textup{eff}}/Z_q^{\textup{eff}}=(\tilde{\omega}_r/\tilde{\omega}_q) (1-s) / (1+s)$. This ratio in the dispersive regime becomes  $\sim (g/\Delta)^{2} + O(g^3)$. Consequently within the dispersive limit the resonator characteristic impedance  is much less than the characteristic impedance associate with transmon.


We  studied a consistent quantization method for superconducting quantum circuits consisting of weakly-anharmonic  transmons coupled to multimode cavities. First we introduced a classical transformation that diagonalizes the interacting harmonic circuit. Once the normal mode basis is identified we quantize the complete circuit Hamiltonian including all anharmonic terms and further simplify it by ignoring nonsecular terms. For the two examples of fundamental importance in quantum computation, i.e. a single transmon coupled to a resonator and two interacting transmons via a bus resonator, we found closed form formula for  dressed frequencies and Kerr coefficients. Our results remain valid and exact for all coupling strengths and detuning frequencies.    This gives a complete  description of the the circuit that otherwise could only be partially achieved by bringing  together various models  each of which valid  within limited domain of parameters.  This indicates that  the black box quantization is a powerful and consistent formalism for studying the physics beyond dispersive regime and scaling up the number of qubits.

\begin{acknowledgments}
We thank David DiVincenzo for many useful discussions. Support from Intelligence Advanced Research Projects Activity (IARPA) under contract W911NF-16-0114 is gratefully acknowledged.
\end{acknowledgments}

\begin{center}
\textbf{\large Supplemental Materials: Exact quantization of superconducting circuits}

%
%
%
\end{center}
\makeatletter
\renewcommand{\theequation}{S\arabic{equation}}
\renewcommand{\thefigure}{S\arabic{figure}}
\renewcommand{\bibnumfmt}[1]{[S#1]}
\renewcommand{\citenumfont}[1]{S#1}

In this supplementary material, we present the details and the derivations of results in the main text. We present the detailed fully quantum mechanical approach to double check our results with a secondary method and come across the same results.


\section{Unitary transformation of canonical variables}
\label{sec.Unitary_SM}
Consider two $N$ dimensional vectors of canonical variables $\mathbf{q}= (q_1, q_2, \cdots, q_N )$ and $\mathbf{p}=(p_1, p_2, \cdots, p_N )$. These variables satisfy the Poisson bracket relation $\{ q_i,p_j\}=\delta_{ij}$ with $i,j=1,2,\cdots,N$ and the definition of $\{f,g\}=\sum_{i=1}^N (\partial f/\partial q_i) (\partial g/\partial p_i) - (\partial f/\partial p_i) (\partial g/\partial q_i)$. 

Let us consider the following unitary transformations takes place on these variables: $Q_i=\sum_{j=1}^N S_{ij} q_j$ and $P_i=\sum_{j=1}^N T_{ij} p_j$.  In order to have the two new variables $\mathbf{Q}$ and $\mathbf{P}$ to be canonical variables they must satisfy similar Poisson bracket relation as those of old variables:  $\{ Q_i,P_j\}=\delta_{ij}$. This indicates that $\{Q_i,P_j\}=\sum_{k=1}^N (\partial Q_i/\partial q_k) (\partial P_j/\partial p_k) - (\partial Q_i/\partial p_k) (\partial P_j/\partial q_k)$. One can easily simplify these relations into: $\sum_{k=1}^N S_{ik} T_{jk} =\delta_{ij}$. Because of the unitarity of the transformation matrices $S$ and $T$ one can see that $\sum_{k=1}^N S_{ik} S_{kj}^\dagger =\delta_{ij}$. For real matrices we have $S^\dagger_{kj}=S_{jk}$, thus $\mathbf{T}=\mathbf{S}$.

%
%


\section{Constraints within exact formula for 2 transmon circuit}
\label{sec.limits_SM}

Another  condition that can be concluded from Eq. (\ref{eq:eignfreq}) in the main text is the following: 
\begin{equation}
{2\sqrt{\frac{f}{3}}\cos \frac{ \textup{cos}^{-1}\left(-\frac{h}{2}\left(\frac{3}{f}\right)^{\frac{3}{2}}\right)-2\pi(k-1)}{3}- \frac{b}{3}}\geq 0  \label{eq:cond_SM}
\end{equation}

By definition we have always $b\leq 0$, therefore the condition can be checked in the cases where $\cos$ function is negative, therefore we need to check   the following condition: $-2\sqrt{f/3}+ |b|/3\geq 0 $, which can be further simplified to ${b^2}/{3}>c$.  Substituting the definitions will introduce the following condition to hold:
\[ \omega_1^4+\omega_2^4 + \omega_3^4 \geq \omega_1^2 \omega_2^2 + \omega_1^2 \omega_3^2 + \omega_2^2 \omega_3^2 - 4 g_1^2 \omega_1 \omega_2 - 4 g_2^2 \omega_2 \omega_3 \]

We take first three terms from right side to the left, then simplify left side to arrive at the following condition: 
\[ (\omega_1^2 - \omega_2^2)^2 + (\omega_2^2 - \omega_3^2)^2 + (\omega_3^2 - \omega_1^2)^2\geq  - 4 g_1^2 \omega_1 \omega_2 - 4 g_2^2 \omega_2 \omega_3 \]
which trivially holds valid without imposing any limitations on parameters.


\section{Unitary transformation for 2 transmons coupled to resonator}
\label{sec. trans 2transmons_SM}
The unitary transformation to diagonal basis in the harmonic sector is carried out by the matrix of normalized eigenstates with columns being eigenvectors, which is 
\begin{equation}
S=\left[\begin{array}{ccc}
\frac{V_1 \gamma_{12}}{N_1} & \frac{V_1 \gamma_{22}}{N_2} & \frac{V_1 \gamma_{32}}{N_3}\\
\frac{V_2 \gamma_{11}}{N_1} & \frac{V_2 \gamma_{21}}{N_2} & \frac{V_2 \gamma_{31}}{N_3}\\
\frac{\gamma_{11} \gamma_{12}}{N_1} & \frac{\gamma_{21} \gamma_{22}}{N_2} & \frac{\gamma_{31} \gamma_{32}}{N_3}
\end{array}\right]\label{eq:Msupp}\end{equation}
with $V_i\equiv g_i\sqrt{4 \omega_r \omega_i}$, $\gamma_{ij}=\bar{\omega}_i^2-\omega_j^2$, and $N_i=\sqrt{V_2^2 \gamma_{i1}^2+V_1^2 \gamma_{i2}^2+\gamma_{i1}^2 \gamma_{i2}^2}$.


\section{Additional interaction terms}
\label{sec_Additional}
In the circuit made of two transmons coupled to a shared resonator, the anharmonic part of Hamiltonian can be simplifed to Eq. (\ref{eq. 2trans}). Below are detailed interaction couplings in terms of bare parameters:
\begin{eqnarray} \nonumber
&& \mathcal{J}_{ik}  = \sum_{j=1,2} \delta_j \left[\frac{1}{3} U_{ji}^{3} U_{jk} + U_{jk} ^{3} U_{ji} + \frac{2}{3} \frac{(U_{ji}U_{jk}U_{j3})^{2} }{U_{j1}U_{j2} } \right], \\ \nonumber
&& \mathcal{K}_{ik} = \sum_{j=1,2} \delta_j U_{ji}^{3}U_{jk} , \ \ \  \mathcal{S}_{ikl} = \frac{4}{3}\sum_{j=1,2} \delta_j \frac{(U_{ji}U_{jk}U_{j3})^{2}}{U_{ji}U_{jk}}, \\
\label{eq. coup_SM}
\end{eqnarray}



\section{Block diagonalization}
\label{sec. SW_SM}
Let us consider the the Hamiltonians of two harmonic oscillators (labeled as 1, 2) coupled to a resonator (labeled as 3):
\begin{eqnarray*} \nonumber
&& H=H_0+\epsilon H_{int},\ \ \ \  H_0\equiv \sum_{i=1,2,3}  \omega_i \hat{\alpha}_i^\dagger \hat{\alpha}_i, \\    \nonumber 
&&   H_{int}\equiv  \sum_{k=1,2} g_k  \left(\hat{\alpha}_{3}\hat{\alpha}_{k}^{\dagger}+\hat{\alpha}_{3}^{\dagger}\hat{\alpha}_{k}\right) \\
  \label{eq. H_SM}
 \end{eqnarray*}

The unperturbed part $H_0$ in the eigenbasis of itself id diagonal, however $H_{int}$ is not. In general we may not be able to find a tranformation to fully diagonal matrix, but instead we can separate out a subset of states from the rest of the states.  The  Schrieffer-Wolff transformation is one way to  block diagonalize the interacting Hamiltonian into  low energy and  high energy sectors.  This usually takes place by transforming the Hamiltonian by the anti-hermitian operator $\exp{S}$ in the following way:  $H_{BD}=\exp{(-S)} H \exp{S}$, which can be expanded into $H_{BD}=\sum_{n=0} [H,S]_{n}/n!$ with $[H,S]_{n+1}=[[H,S]_{n},S]$ and $[H,S]_{0}=H$.  One can in principle assume a geometric series expansion of the transformation matrix:   $S=\sum_{i=0} (\epsilon )^i S_i$; however  given that the zeroth or order of $H_{BD}$ is $H_{{BD}_0}=[H_0,S_0]=H_0$ therefore $S_0$ must be diagonal too which is in fact inconsistent with the definition of $S$ to be anti-hermitian and block-off-diagonal, therefore always $S_0=0$.  In the first order the Hamiltonian is already given by $H_{int}$ which can be made of block-diagonal (bd) and block-off-diagonal (bod) matrices $H_{int}=H_{int}^{\textup{bd}}+H_{int}^{\textup{bod}}$ . Therefore $H_{BD_1}=[H_0,S_1]=-H_{int}^{\textup{bod}}$. In the second order: $H_{BD_2}=[H_0,S_2]+[H_{int},S_1]+(1/2)[[H_0,S_1],S_1]$, and so on.  Putting all together one can find the effective Hamiltonian up to the second order $H_{BD}=H_0+H_{int}^{\textup{bd}}+(1/2)[H_{int}^{\textup{bod}},S_1]$. 

Using the relations above for the Hamiltonian of Eq. (\ref{eq. H_SM}) in which the interaction is block-off-diagonal one can use the following ansatz 

\begin{equation}
S_1=- \sum_{k=1,2} g_{k}\left(\hat{\alpha_{3}}\hat{\alpha'}_{k}^{\dagger}-\hat{\alpha}_{3}^{\dagger}\hat{\alpha'}_{k}\right).
\end{equation} 
with $\hat{\alpha'}_{k}\equiv \sum_{n=0}^\infty \sqrt{n+1}(\omega_3-\omega_k)^{-1} |n\rangle \langle n+1|$ being the modified ladder operator for $k$-th transmon, given that the normal ladder operator for the same transmon is $\hat{\alpha}_{k}\equiv \sum_{n=0}^\infty \sqrt{n+1} |n\rangle \langle n+1|$ . 

One can explicitly determine the effective Hamiltonian up to the second order of perturbation theory  becomes
\begin{equation}
H_{BD}=H_0-\sum_{i,j=1,2; (i\neq j)} \frac{g_i g_j}{2} \left(\hat{\alpha_{i}}\hat{\alpha'}_{j}^{\dagger}+\hat{\alpha}_{i}^{\dagger}\hat{\alpha'}_{j}\right)
\label{eq.HBD_SM}
\end{equation}


\section{Resonant transmons}
\label{sec. resonance_SM}

In a circuit with two transmons in resonance $\omega_1=\omega_2\equiv \omega$ and homogeneous coupling and anharmonicity  $g_1=g_2\equiv g$  and $\delta_1=\delta_2\equiv \delta$ the harmonic Hamiltonian is $H_{\textup{har.}}=\frac{1}{2} \omega^2 (Q_{1}^{2}+Q_2^2)+\frac{1}{2} \omega_r^2 Q_{r}^{2}+\frac{1}{2} (X_1^2+X_2^2+X_{r}^{2})+ g \sqrt{4\omega \omega_{r}}(Q_{1}+Q_2)Q_{3}$.  Defining the vectors $\mathbf{Q}=(Q_1, Q_2, Q_r)^T$ and $\mathbf{P}=(P_1, P_2, P_r)^T$, this Hamiltonian can be rewritten as $H_{\textup{har.}}=\frac{1}{2}\mathbf{Q}^\textup{T} \mathbf{M Q}+\frac{1}{2}\mathbf{X}^\textup{T}\mathbf{  X}$  with the matrix $\mathbf{M}$ being 
\begin{equation}
\mathbf{M}=\left[\begin{array}{ccc}
\omega^{2} & 0 & V\\
0 & \omega^{2} & V\\
V& V & \omega_{r}^{2}
\end{array}\right]\label{eq:M3dar3_SM}
\end{equation}
with $V\equiv g\sqrt{4\omega\omega_{r}}$. 
Because the off diagonal elements are identical,  it is easy to find the eigenvalues, which are
\[  \omega,\ \ \  \sqrt{\frac{\omega^2+\omega_r^2 \pm \sqrt{(\omega^2-\omega_r^2)^2+8V^2}}{2}} \]
At the exterm resonance with $\omega_r=\omega$ the eigenenergies will become
\[   \omega,\ \ \  \omega \sqrt{1 \pm  \frac{2\sqrt{2} g }{\omega}} \]

In the limit of small coupling $g\ll \omega$ this can be simplified to 
\[   \omega,\ \ \   \omega \pm  \sqrt{2} g  
\]
\


\section{Anaharmonicity}
\label{sec. anharmonicity_SM}

Consider the following Bogoliubov transformations for transmon ladder operator:
\begin{equation}
\hat{a}_{n}  =  \sum_{m}A_{nm}\hat{\alpha}_{m}+B_{nm}\hat{\alpha}_{m}^{\dagger}
\label{eq:bog}
\end{equation}
and using the relation between transmon charge number and phase and the ladder operator $\hat{a}_{n}=\sqrt{\frac{\omega_{n}}{2}}\hat{q}_{n}+i\sqrt{\frac{1}{2\omega_{n}}}\hat{p}_{n}$, and its conjugate as well as similar in the transformed basis $\hat{\alpha}_{n}=\sqrt{\frac{\tilde{\omega}_{n}}{2}}\hat{Q}_{n}+i\sqrt{\frac{1}{2\tilde{\omega}_{n}}}\hat{P}_{n}$, one can find
\begin{eqnarray*}
&& A_{nm}  =  \left(\sqrt{\frac{\omega_{n}}{8\tilde{\omega}_{m}}}+\sqrt{\frac{\tilde{\omega}_{m}}{8\omega_{n}}}\right)S_{nm},\\ 
& & B_{nm}  =  \left(\sqrt{\frac{\omega_{n}}{8\tilde{\omega}_{m}}}-\sqrt{\frac{\tilde{\omega}_{m}}{8\omega_{n}}}\right)S_{nm}
\end{eqnarray*}
in which $\tilde{\omega}$ is the frequency in the transformed basis. 


The anharmonicity in Hamiltonian will be $-\frac{\delta_{i}}{12}\left(a_{i}-a_{i}^{\dagger}\right)^{4}$. The operator part can be Bogoliubov transformed to the new basis, keeping terms with as many creations as annihilations, ignoring frequencies: 
\begin{widetext}
\begin{eqnarray*}
&& \left(a_{n}-a_{n}^{\dagger}\right)^{4}  = \\ && 6\sum_{m=1}^{3}\left(A_{nm}-B_{nm}\right)^{4}\left[\left(\hat{\alpha}_{m}^{\dagger}\hat{\alpha}_{m}\right)^{2}+\hat{\alpha}_{m}^{\dagger}\hat{\alpha}_{m}\right]\\
 &  & +6\sum_{m<k}\left(A_{nm}-B_{nm}\right)^{2}\left(A_{nk}-B_{nk}\right)^{2}\left[\hat{\alpha}_{m}^{2}\hat{\alpha}_{k}^{\dagger2}+\hat{\alpha}_{m}^{\dagger2}\hat{\alpha}_{k}^{2}+4\hat{\alpha}_{m}^{\dagger}\hat{\alpha}_{m}\hat{\alpha}_{k}^{\dagger}\hat{\alpha}_{k}+2\hat{\alpha}_{m}^{\dagger}\hat{\alpha}_{m}+2\hat{\alpha}_{k}^{\dagger}\hat{\alpha}_{k}\right]\\
 &  & +4\sum_{m\neq k}\left(A_{nm}-B_{nm}\right)^{3}\left(A_{nk}-B_{nk}\right)\left(\hat{\alpha}_{m}^{2}\hat{\alpha}_{m}^{\dagger}\hat{\alpha}_{k}^{\dagger}+\hat{\alpha}_{m}^{\dagger2}\hat{\alpha_{m}}\hat{\alpha}_{k}+2\hat{\alpha}_{m}^{\dagger}\hat{\alpha}_{m}\hat{\alpha_{m}}\hat{\alpha}_{k}^{\dagger}+2\hat{\alpha}_{m}^{\dagger}\hat{\alpha}_{m}\hat{\alpha}_{m}^{\dagger}\hat{\alpha}_{k}+\hat{\alpha_{m}}\hat{\alpha}_{k}^{\dagger}+\hat{\alpha}_{m}^{\dagger}\hat{\alpha}_{k}\right)\\
 &  & +8\sum_{m\neq k\neq l}\left(A_{nm}-B_{nm}\right)^{2}\left(A_{nk}-B_{nk}\right)\left(A_{nl}-B_{nl}\right)\left(\hat{\alpha}_{m}^{2}\hat{\alpha}_{l}^{\dagger}\hat{\alpha}_{k}^{\dagger}+\hat{\alpha}_{m}^{\dagger2}\hat{\alpha_{l}}\hat{\alpha}_{k}+2\hat{\alpha}_{m}^{\dagger}\hat{\alpha}_{m}\hat{\alpha_{l}}\hat{\alpha}_{k}^{\dagger}+2\hat{\alpha}_{m}^{\dagger}\hat{\alpha}_{m}\hat{\alpha}_{l}^{\dagger}\hat{\alpha}_{k}+\hat{\alpha_{l}}\hat{\alpha}_{k}^{\dagger}+\hat{\alpha}_{l}^{\dagger}\hat{\alpha}_{k}\right)
\end{eqnarray*}
\end{widetext}


\section{Bogoliubov transformation for Hamiltonian diagonalization}
\label{sec.Bogoliuboc_SM}

In this section we use quantum Hamiltonian of a transmon coupled to a resonator is $H  =  4E_{c}n-E_{J}\cos\phi+H_\textup{res}$. Separating the harmonic sector and the anharmonic sector, and using Bogoliubov transformation we diagonalize the interacting harmonic sector into a diagonal quantum harmonic Hamiltonian. We find all Bogoliubov transformation coefficients, which turns out to be similar to the results we took from semiclassical analysis.  

Given that charge number operator is proportional to ladder operators $n  \sim  2^{-\frac{1}{4}}\left(a+a^{\dagger}\right)$ and phase is the conjugate variable $\phi\sim2^{\frac{1}{4}}\left(a-a^{\dagger}\right)$, and the resonator Hamiltonian is $H_\textup{res}=\omega_{r}b^{\dagger}b$, the circuit Hamiltonian can be written as $H  =  \omega_{q}a^{\dagger}a-\frac{\delta}{12}\left(a-a^{\dagger}\right)^{4}+\omega_{r}b^{\dagger}b+g\left(a+a^{\dagger}\right)\left(b+b^{\dagger}\right)$ with harmonic part being $H_\text{har}  = \omega_{q}a^{\dagger}a+\omega_{r}b^{\dagger}b+g\left(a+a^{\dagger}\right)\left(b+b^{\dagger}\right)$.  

We would like to Bogoliubov-transform the Hamiltonian into a diagonal Hamiltonian $\mathcal{H} $: 
\[\mathcal{H}  =  \tilde{\omega}_{q}\alpha^{\dagger}\alpha+\tilde{\omega}_{r}\beta^{\dagger}\beta-\frac{1}{12}\left({\chi}_{q}^{\frac{1}{4}}\left(\alpha-\alpha^{\dagger}\right)+{\chi}_{r}^{\frac{1}{4}}\left(\beta-\beta^{\dagger}\right)\right)^{4}\]

We use a technique widely used in second quantized QFT, which is to
Bogoliubov-transofmation creation and annihilation operators
\[ \hat{a}  =  A\hat{\alpha}+B\hat{\beta}+C\alpha^{\dagger}+D\beta^{\dagger}, \ \ \ \ \hat{b}  =  E\hat{\alpha}+F\hat{\beta}+G\alpha^{\dagger}+H\beta^{\dagger}\]

Eight equations are needed to determines coefficients; four by enforcing
that transformed Hamiltonian preserves eigenvalues, which is equivalent
to equating $H_{ho}$ and $\mathcal{H}_{ho}$ and setting coefficients
of $\hat{\alpha}\hat{\alpha},\hat{\beta}\hat{\beta},\hat{\alpha}\hat{\beta}$
and $\hat{\alpha}\hat{\beta}^{\dagger}$to zero, respectively:
\begin{eqnarray}
 &  & \omega_{q}AC^{*}+\omega_{r}EG^{*}+g\left(A+C^{*}\right)\left(E+G^{*}\right)=0\label{eq:1-1}\\
 &  & \omega_{q}BD^{*}+\omega_{r}FH^{*}+g\left(B+D^{*}\right)\left(F+H^{*}\right)=0\label{eq:2-1}\\
 &  & \omega_{q}\left(BC^{*}+AD^{*}\right)+\omega_{r}\left(FG^{*}+EH^{*}\right)\label{eq:3-1}\\
 &  & \quad\quad+g\left[\left(A+C^{*}\right)\left(F+H^{*}\right)+\left(B+D^{*}\right)\left(E+G^{*}\right)\right]=0\nonumber \\
 &  & \omega_{q}\left(DC^{*}+AB^{*}\right)+\omega_{r}\left(HG^{*}+EF^{*}\right)\label{eq:4-1}\\
 &  & \quad\quad+g\left[\left(A+C^{*}\right)\left(F+H^{*}\right)+\left(B+D^{*}\right)\left(E+G^{*}\right)\right]=0\nonumber 
\end{eqnarray}
The other four are determined by enforcing commutation relations,
i.e. $[a,a^{\dagger}]=[b,b^{\dagger}]=1$ and $[a,b]=[a,b^{\dagger}]=0$,
respectively, given that $[\alpha,\alpha^{\dagger}]=[\beta,\beta^{\dagger}]=1$
and zero otherwise:
\begin{eqnarray}
|A|^{2}+|B|^{2}-|C|^{2}-|D|^{2} & = & 1,\label{eq:5-1}\\
|E|^{2}+|F|^{2}-|G|^{2}-|H|^{2} & = & 1,\label{eq:6-1}\\
AG+BH-CE-DF & = & 0,\label{eq:7-1}\\
AE^{*}+BF^{*}-CG^{*}-DH^{*} & = & 0.\label{eq:8-1}
\end{eqnarray}

For simplicity we assume coefficients are real-valued, but the equations are difficult to be analytically solved. A practical simplification can be achieved by defining new variables
\[ A_{\pm}\equiv A\pm C,\;B_{\pm}\equiv B\pm D,\;E_{\pm}\equiv E\pm G,\;F_{\pm}\equiv F\pm H \]
 which reformulates equations given above to the followings:
\begin{eqnarray*}
 &  & \omega_{q}\left(A_{+}^{2}-A_{-}^{2}\right)+\omega_{r}\left(E_{+}^{2}-E_{-}^{2}\right)+4gE_{+}A_{+}=0,\label{eq:9-1}\\
 &  & \omega_{q}\left(B_{+}^{2}-B_{-}^{2}\right)+\omega_{r}\left(F_{+}^{2}-F_{-}^{2}\right)+4gF_{+}B_{+}=0,\label{eq:10-1}\\
 &  & \omega_{q}\left(A_{+}B_{+}-A_{-}B_{-}\right)+\omega_{r}\left(E_{+}F_{+}-E_{-}F_{-}\right)\nonumber \\
 &  & \qquad\qquad\qquad\qquad+2g\left(A_{+}F_{+}+B_{+}E_{+}\right)=0,\label{eq:11-1}\\
 &  & \omega_{q}A_{-}B_{-}+\omega_{r}E_{-}F_{-}=0,\label{eq:12-1}\\
 &  & A_{-}A_{+}+B_{+}B_{-}=1,\label{eq:13-1}\\
 &  & E_{-}E_{+}+F_{+}F_{-}=1,\label{eq:14-1}\\
 &  & A_{-}E_{+}+B_{-}F_{+}=0,\label{eq:15-1}\\
 &  & A_{+}E_{-}+B_{+}F_{-}=0.\label{eq:16-1}
\end{eqnarray*}

Given that one may solve the Bogoliubov coefficient equtions, we can
determine new frequencies in $\mathcal{H}$:
\begin{eqnarray*}
 &  & \bar{\omega}_{r}=\frac{\omega_{q}}{2}\left(B_{+}^{2}+B_{-}^{2}\right)+\frac{\omega_{r}}{2}\left(F_{+}^{2}+F_{-}^{2}\right)+2gB_{+}F_{+}\label{eq:wr-1}\\
 &  & \bar{\omega}_{q}=\frac{\omega_{q}}{2}\left(A_{+}^{2}+A_{-}^{2}\right)+\frac{\omega_{r}}{2}\left(E_{+}^{2}+E_{-}^{2}\right)+2gA_{+}E_{+}\label{eq:wq-1}
\end{eqnarray*}

One can easily prove that $F_{+}F_{-}=A_{+}A_{-}$, which simplifies
equations and helps to find the following two important equalities: 
\begin{eqnarray*}
E_{+}^{2} & = & \frac{\omega_{r}A_{+}\left(1-A_{-}A_{+}\right)}{\omega_{q}A_{-}},\qquad E_{-}^{2}=\frac{\omega_{q}A_{-}\left(1-A_{-}A_{+}\right)}{\omega_{r}A_{+}}
\end{eqnarray*}

Substituting them in Eq. (\ref{eq:9-1}) we find one equation between
$A_{\pm}$:
\begin{eqnarray}
&& \left[\omega_{q}^{2}A_{-}\left(A_{+}^{3}-A_{-}\right)+\omega_{r}^{2}A_{+}^{2}\left(1-A_{+}A_{-}\right)\right]^{2}\nonumber \\
&& -16\omega_{r}\omega_{q}g^{2}A_{+}^{5}A_{-}\left(1-A_{-}A_{+}\right)  =  0\label{eq:meq1-1}
\end{eqnarray}

This is one of the main equations we need to solve. Another one can
be determined taking some non-trivial steps listed below: We use Eq.
(\ref{eq:11-1}), substitute $B_{\pm}$ from Eqs. (\ref{eq:15-1},\ref{eq:16-1}),
multiply two side in $E_{+}F_{-}^{2}F_{+}$ and simplify it, magically
the final equation is again a second equation that relation $A_{\pm}$:
\begin{eqnarray}
\left(1-A_{-}A_{+}\right)A_{+}A_{-}\left(\frac{\omega_{r}^{2}}{2\omega_{q}}-\frac{\omega_{q}}{2}\right)^{2}\nonumber \\
-\left(2A_{-}A_{+}-1\right)^{2} & = & 0\label{eq:meq2-1}
\end{eqnarray}

Now we solve these two equations together. To do so we first define
$x=A_{+}A_{-}$ and substitute in Eq. (\ref{eq:meq2-1}): $a(1-x)x-(2x-1)^{2}=0$
with $a\equiv\frac{\Delta^{2}\Sigma^{2}}{4g^{2}\omega_{r}\omega_{q}}$ and $\Sigma=\omega_{r}+\omega_{q}$ and $\Delta=\omega_{r}-\omega_{q}$.
Exact real-valued solution is
\[ A_{-}A_{+}=\frac{1}{2}+\frac{1}{2}s,\qquad s^{-1}\equiv\sqrt{1+\frac{16g^{2}\omega_{r}\omega_{q}}{\Delta^{2}\Sigma^{2}}}
\]
and substituting in Eq. (\ref{eq:meq1-1}) determines exact real-valued
$A_{\pm}$:
\begin{eqnarray*}
A_{-} & = & 2^{-\frac{3}{4}}\omega_{q}^{-\frac{1}{2}}\sqrt{1+s}\left(\omega_{q}^{2}+\omega_{r}^{2}-\Delta\Sigma s^{-1}\right)^{\frac{1}{4}}\\
A_{+} & = & 2^{-\frac{1}{4}}\omega_{q}^{\frac{1}{2}}\sqrt{1+s}\left(\omega_{q}^{2}+\omega_{r}^{2}-\Delta\Sigma s^{-1}\right)^{-\frac{1}{4}}\\
E_{-} & = & -2^{-\frac{3}{4}}\omega_{r}^{-\frac{1}{2}}\sqrt{1-s}\left(\omega_{q}^{2}+\omega_{r}^{2}-\Delta\Sigma s^{-1}\right)^{\frac{1}{4}}\\
E_{+} & = & -2^{-\frac{1}{4}}\omega_{r}^{\frac{1}{2}}\sqrt{1-s}\left(\omega_{q}^{2}+\omega_{r}^{2}-\Delta\Sigma s^{-1}\right)^{-\frac{1}{4}}\\
F_{-} & = & 2^{-\frac{3}{4}}\omega_{r}^{-\frac{1}{2}}\sqrt{1+s}\left(\omega_{q}^{2}+\omega_{r}^{2}+\Delta\Sigma s^{-1}\right)^{\frac{1}{4}}\\
F_{+} & = & 2^{-\frac{1}{4}}\omega_{r}^{\frac{1}{2}}\sqrt{1+s}\left(\omega_{q}^{2}+\omega_{r}^{2}+\Delta\Sigma s^{-1}\right)^{-\frac{1}{4}}\\
B_{-} & = & 2^{-\frac{3}{4}}\omega_{q}^{-\frac{1}{2}}\sqrt{1-s}\left(\omega_{q}^{2}+\omega_{r}^{2}+\Delta\Sigma s^{-1}\right)^{\frac{1}{4}}\\
B_{+} & = & 2^{-\frac{1}{4}}\omega_{q}^{\frac{1}{2}}\sqrt{1-s}\left(\omega_{q}^{2}+\omega_{r}^{2}+\Delta\Sigma s^{-1}\right)^{-\frac{1}{4}}
\end{eqnarray*}

In order to find $F_{\pm}$ yet we need to simplify Eq. (\ref{eq:10-1})
by multiplying on both sides on $F_{-}F_{+}$ and rewriting $B_{\pm}$
in terms of $A_{\pm}$, $E_{\pm}$ and $F_{\pm}$ as shown in Eqs.
(\ref{eq:15-1},\ref{eq:16-1}):
\[ \left(\frac{F_{-}}{F_{+}}\right)^{2}=\frac{1}{2}\frac{\omega_{q}^{2}+\omega_{r}^{2}+\Delta\Sigma s^{-1}}{\omega_{r}^{2}}
\]

Defining 
\begin{eqnarray*}
K_{\pm} & \equiv & 2^{-\frac{1}{4}}\left(\omega_{q}^{2}+\omega_{r}^{2}\pm\Delta\Sigma s^{-1}\right)^{\frac{1}{4}}
\end{eqnarray*}

then
\begin{eqnarray*}
 &  & A=\frac{\sqrt{1+s}}{2^{3/2}}\left(\frac{\sqrt{\omega_{q}}}{K_{-}}+\frac{K_{-}}{\sqrt{\omega_{q}}}\right), \\ 
  &  & B=\frac{\sqrt{1-s}}{2^{3/2}}\left(\frac{\sqrt{\omega_{q}}}{K_{+}}+\frac{K_{+}}{\sqrt{\omega_{q}}}\right),\\ 
   & & C=\frac{\sqrt{1+s}}{2^{3/2}}\left(\frac{\sqrt{\omega_{q}}}{K_{-}}-\frac{K_{-}}{\sqrt{\omega_{q}}}\right)\\
& & D=\frac{\sqrt{1-s}}{2^{3/2}}\left(\frac{\sqrt{\omega_{q}}}{K_{+}}-\frac{K_{+}}{\sqrt{\omega_{q}}}\right)\\
 &  & E=\frac{-\sqrt{1-s}}{2^{3/2}}\left(\frac{\sqrt{\omega_{r}}}{K_{-}}+\frac{K_{-}}{\sqrt{\omega_{r}}}\right),\\
 &  & F=\frac{\sqrt{1+s}}{2^{3/2}}\left(\frac{\sqrt{\omega_{r}}}{K_{+}}+\frac{K_{+}}{\sqrt{\omega_{r}}}\right),\\
 & & G=\frac{-\sqrt{1-s}}{2^{3/2}}\left(\frac{\sqrt{\omega_{r}}}{K_{-}}-\frac{K_{-}}{\sqrt{\omega_{r}}}\right)\\
 & & H=\frac{\sqrt{1+s}}{2^{3/2}}\left(\frac{\sqrt{\omega_{r}}}{K_{+}}-\frac{K_{+}}{\sqrt{\omega_{r}}}\right)
\end{eqnarray*}

We can expand the functions in terms of small coupling $g$ to any
order. Below are results up to the fourth order:

Substituting in definition of new frequencies one finds: 
\begin{eqnarray*}
\tilde{\omega}_{r} & = & \left(2s\right)^{-\frac{1}{2}}\sqrt{\left(\omega_{q}^{2}+\omega_{r}^{2}\right)s+\Delta\Sigma}\\
\tilde{\omega}_{q} & = & \left(2s\right)^{-\frac{1}{2}}\sqrt{\left(\omega_{q}^{2}+\omega_{r}^{2}\right)s-\Delta\Sigma}
\end{eqnarray*}

In the weak interaction limit these frequecies turn into Lamb and
Stark shifts. Below we evaluate them up to fourth order: 
\begin{eqnarray*}
\tilde{\omega}_{r} & = & \omega_{r}+\frac{2\omega_{q}g^{2}}{\Delta\Sigma}-\frac{2g^{4}\omega_{q}^{2}\left(5\omega_{r}^{2}-\omega_{q}^{2}\right)}{\omega_{r}\Delta^{3}\Sigma^{3}}+O\left(g^{5}\right)\\
\tilde{\omega}_{q} & = & \omega_{q}-\frac{2g^{2}\omega_{r}}{\Delta\Sigma}-\frac{2g^{4}\omega_{r}^{2}\left(\omega_{r}^{2}-5\omega_{q}^{2}\right)}{\omega_{q}\Delta^{3}\Sigma{}^{3}}+O\left(g^{5}\right)
\end{eqnarray*}

Anharmonicity can be easily derived using the following relation:
\begin{eqnarray*}
\left(a-a^{\dagger}\right)^{4} & = & 6\left(A-C\right)^{4}\left(\left(\alpha^{\dagger}\alpha\right)^{2}+\alpha^{\dagger}\alpha\right)\\
 &  & +6\left(B-D\right)^{4}\left(\left(\beta^{\dagger}\beta\right)^{2}+\beta^{\dagger}\beta\right)\\
 &  & +12\left(A-C\right)^{2}\left(B-D\right)^{2}\left(2\alpha^{\dagger}\alpha\beta^{\dagger}\beta+\alpha^{\dagger}\alpha+\beta^{\dagger}\beta\right)
\end{eqnarray*}

\end{document}